\documentclass[sigconf]{acmart}
\usepackage{multirow}
\usepackage{xcolor}
\usepackage{color, colortbl}
\usepackage{footmisc}
\usepackage{float}

\usepackage{booktabs} % For formal tables
\usepackage[
  moderate
]{savetrees}
\usepackage{blindtext}

\setcopyright{rightsretained}
\newcommand{\aha}[1]{\textcolor{blue}{\textbf{*AHA*}: #1}}

\newcommand{\bs}[1]{\textcolor{purple}{\textbf{*B.S*}: #1}}

\acmDOI{}

\acmConference[WSDM'19]{ACM WSDM conference}{February 2019}{Melbourne, Australia}
\acmYear{2019}
\copyrightyear{2019}

\acmPrice{15.00}

\begin{document}
\title{Characterizing and Predicting Email Deferral Behavior}

\begin{comment}
\author{Bahareh Sarrafzadeh\inst{1} \and
Ahmed Hassan Awadallah\inst{2} \and 
Christopher H. Lin\inst{2} \and
Chia-Jung Lee\inst{2} \and
Milad Shokouhi\inst{2} \and
Susan T. Dumais\inst{2}
}

\institution{University of Waterloo, Waterloo, ON, Canada \\
\email{bsarrfaz@uwaterloo.ca} \and
Microsoft, Redmond, WA, USA \\
\email{\{hassanam,christol,cjlee,milads,sdumais\}@microsoft.com}
}

\end{comment}

\author{Bahareh Sarrafzadeh}
\authornote{Work done while at Microsoft.}
\affiliation{%
  \institution{University of Waterloo}
  %\streetaddress{One Microsoft Way}
  \city{Waterloo}
  \state{Canada}
}
\email{bsarrafz@uwaterloo.ca}

\author{Ahmed Hassan Awadallah, Christopher H. Lin, Chia-Jung Lee, Milad Shokouhi, Susan T. Dumais}
%\authornote{Dr.~Trovato insisted his name be first.}
\affiliation{%
  \institution{Microsoft}
  \streetaddress{One Microsoft Way}
  \city{Redmond}
  \state{WA, USA}
}
\email{{hassanam,christol,cjlee,milads,sdumais}@microsoft.com}

\renewcommand{\shortauthors}{Sarrafzadeh, B., H. Awadallah, A., Lee, CJ., Lin, Ch., Shokouhi, M., Dumais, S.} 

\fancyhead{}
\begin{comment}

\author{Christopher Lin}
%\authornote{Dr.~Trovato insisted his name be first.}
\affiliation{%
  \institution{Microsoft Research}
  \streetaddress{One Microsoft Way}
  \city{Redmond}
  \state{WA, USA}
}
\email{Christopher.Lin@microsoft.com}

\author{Chia-Jung Lee}
%\authornote{Dr.~Trovato insisted his name be first.}
\affiliation{%
  \institution{Microsoft Research}
  \streetaddress{One Microsoft Way}
  \city{Redmond}
  \state{WA, USA}
}
\email{email@microsoft.com}

\author{Milad Shokouhi}
%\authornote{Dr.~Trovato insisted his name be first.}
\affiliation{%
  \institution{Microsoft Research}
  \streetaddress{One Microsoft Way}
  \city{Redmond}
  \state{WA, USA}
}
\email{milads@microsoft.com}

\author{Susan T. Dumais}
%\authornote{Dr.~Trovato insisted his name be first.}
\affiliation{%
  \institution{Microsoft Research}
  \streetaddress{One Microsoft Way}
  \city{Redmond, USA}
  \state{WA}
}
\email{sdumais@microsoft.com}

% \authornote{The secretary disavows any knowledge of this author's actions.}
% \affiliation{%

% \renewcommand{\shortauthors}{B. Trovato et al.}

\end{comment}

\begin{abstract}
Email triage involves going through \textit{unhandled} emails and deciding what to do with them. This familiar process can become increasingly challenging as the number of unhandled email grows. %\sd{I'm ok with keeping this definition of triage, but I'm not sure why triage involves only unhandled messages}. 
%\bs{You are right. Neustaedter et al. (2005) found that triaging is primarily performed on unread emails in the inbox, with some people also looking at read items. So I remember we did have the same discussion regarding how triage should be defined. Triaging emails that haven't been looked at before is closer to the process of triage (in medical domain) where a decision about *when* that patient (or email) should be handled is made. Yet it's not clear if we should consider an email 'handled' when it's looked at but was left for later to be dealt with (i.e. deferral case).} 
During a triage session, users commonly \textit{defer} handling emails that they cannot immediately deal with to later. These deferred emails, are often related to tasks that are postponed until the user has more time or the right information to deal with them. In this paper, through qualitative interviews and a large-scale log analysis, we study when and what \textit{enterprise} email users tend to defer. We found that users are more likely to defer emails when handling them involves replying, reading carefully, or clicking on links and attachments. We also learned that the decision to defer emails depends on many factors such as user's workload and the importance of the sender.

Our qualitative results suggested that deferring is very common, and our quantitative log analysis confirms that 12\% of triage sessions and 16\% of daily active users had at least one deferred email on weekdays. We also discuss several \textit{deferral strategies} such as marking emails as unread and flagging that are reported by our interviewees, and illustrate how such patterns can be also observed in user logs.

Inspired by the characteristics of deferred emails and contextual factors involved in deciding if an email should be deferred, we train a classifier for predicting whether a recently triaged email is actually deferred. Our experimental results suggests that deferral can be classified with modest effectiveness. Overall, our work provides novel insights about how users handle their emails and how deferral can be modeled.%, and how email clients can be redesigned to support deferral better. 
%\ms{we should make the abstract shorter}
%\cl{I think we need to say why we care about deferral - what is the end goal. Right now it reads like we just did a bunch of things about deferral without any reason. Why is it important to characterize and predict deferral?}

%until later to manage overflow: emails that cannot or should not be dealt with now.
\end{abstract}

\vspace{-0.3mm}
\keywords{Email Management, Triage, Deferral, Users Behavior Modeling}
\vspace{-0.3mm}
\copyrightyear{2019} 
\acmYear{2019} 
\setcopyright{acmcopyright}
\acmConference[WSDM '19]{The Twelfth ACM International Conference on Web Search and Data Mining}{February 11--15, 2019}{Melbourne, VIC, Australia}
\acmBooktitle{The Twelfth ACM International Conference on Web Search and Data Mining (WSDM '19), February 11--15, 2019, Melbourne, VIC, Australia}
\acmPrice{15.00}
\acmDOI{10.1145/3289600.3291028}
\acmISBN{978-1-4503-5940-5/19/02}

% Use the following line to eliminate the ACM reference format for the work.
%\settopmatter{printacmref=false} 
\maketitle

\section{Introduction}

% General problem
Email is one of the most popular online activities and remains a major tool for communication and collaboration. It is estimated that 269 billion emails were sent and received per day in 2017~\cite{Radicati2017}, and %Email is particularly popular for work related communications as 
studies show that information workers tend to spend up to 28\% of their time reading and answering email~\cite{McKinsey2012}. 
%\bs{There is a disconnect between why studying email is important and why we are studying Triage / deferral in this paper. The missing link is the concept of Email Overload and the Task Management aspect of emails.}

% email overload and triage
Email usage has significantly evolved beyond \textit{communication} to encompass other areas like \textit{task management}, \textit{archiving}, etc. \citet{Mackay:1988:MJC:62266.62293} presented one of the earliest studies examining how email was being used for more than just communication, a phenomenon often referred to as ``email overload'' \cite{Whittaker}. Since then many researchers \cite{Whittaker,Bellotti:2003:TET:642611.642672,Ducheneaut:2001:EHE:382899.383305,Gwizdka2} have studied the close tie between people's tasks and their email practices.  \citet{Venolia2001} then consolidated these findings into five areas of email activity: flow, triage, task management, archive, and retrieve.
 
%introduce triage and deferral
Email \textit{triage} is the process of going through unhandled email and deciding what to do with it. Email triage can quickly become a serious problem for users as the number of unhandled emails grows. During a Triage session, users commonly defer emails until later to manage overflow\cite{Siu:2006}. Email \textit{deferral} is directly related to task management, and occurs because people have insufficient time to take an immediate action  or they need to gather information before they can act on a message ~\cite{Bellotti:2003:TET:642611.642672,Whittaker2}.
Dabbish et al.~\cite{Dabbish} showed that people defer responding to 37\% of messages that need a reply. Similarly, in a sample of the logs of a popular email client we analyzed, we found that while around 10\% of all messages receive a Reply, a ReplyAll or a Forward action; 26\% of these actions are taken at a later time (not immediately following the first read) indicating the significance of deferral. 

%Email \textit{triage} is the process of going through unhandled email and deciding what to do with it. Email triage can quickly become a serious problem for users as the number of unhandled email grows. During a Triage session, users commonly defer emails until later to manage overflow: emails that cannot or should not be dealt with now.  \ms{this paragraph looks misplaced, we should move or remove.}

%\textbf{Why do people defer acting on emails?}\\
%Previous research has shown that users frequently deferred messages  as a strategy for managing their attention budget since only the most important emails get handled during early scans \ms{cite}.  
The fact that a user defers an email does not imply that the message is less important. A deferred email could be very important and therefore requires careful examination and a well crafted reply. Alternatively, it could be not important enough to warrant immediate attention. %Furthermore, it may contain useful reference information that the user decides to consume later when more time is available.
Deferral could also be a result of other factors unrelated to the message such as the current user workload and the device she is currently using. For example, a common scenario for deferral involves the increasing use of mobile devices for day-to-day task management. Not only does triage play a more prominent role on mobile devices, users also need to accomplish it more quickly because of the short, intermittent nature of mobile interactions~\cite{matthews2009no}. Past research on smart phone use suggests that mobile users primarily identify what emails to delete and to handle immediately, and defer handling most messages until they reach a larger device ~\cite{pierce2010triage}.

%What are we doing and %how are we doing it
Understanding email deferral characteristics, strategies and motivations could help develop new experiences to empower people to get tasks done more efficiently. In this paper, we present a detailed study of email deferral. We employ a \textit{mixed methods} approach where we combine qualitative and quantitative analysis to better characterize deferral in email.  We present a large-scale log analysis of forty thousand anonymized users of a popular commercial email client. We complement the log analysis with a qualitative study where we perform interviews to gain more insights into the motivation for the observed user behavior. Inspired by the insights we develop from characterizing email deferral, we define a prediction task to support email deferral using features of the email message, user workload and behavioral logs. We also discuss the implications of this work on designing email clients to help users be more efficient with managing their mailboxes.

%contributions
Previous research has mostly focused on email organization and management \cite{Venolia2001,Whittaker2,Siu:2006}. However, to the best of our knowledge, understanding why people defer emails and what strategies they use to go back to them have received much less attention. In particular, we focus on the following research questions:
%\begin{itemize}
%\item 
How is deferral defined and perceived by users? [\textbf{RQ1}]
%\item 
How common is deferral? [\textbf{RQ2}]
%\item 
What motivations are behind deferral? [\textbf{RQ3}]
%\item 
What factors impact deferral for an email? [\textbf{RQ4}]
%\item 
What strategies are employed for deferral? [\textbf{RQ5}]
%\end{itemize}

\begin{comment}
%The main contributions of this work  can be summarized as follows:
%\begin{enumerate}
%\item 
To answer these questions, we make the following contributions: (1) We begin with a qualitative study using interviews to understand why and how people defer messages and what affects their decision.
%\item 
(2) We present a detailed analysis of email deferral by performing a large-scale log study of email interactions of thousands of users that largely supports the trends observed in our qualitative interviews. 
%\item 
(3) Inspired by our qualitative and quantitative learnings, we define the task of predicting email deferral from user interaction logs and describe a model for identifying deferred messages using a large number of features. 
%\item 
(4) We discuss the implications of our study on designing email clients. %Finally, we provide an overview of the most related work in Section~\ref{sec-relatedwork}, before we conclude the paper in Section~\ref{sec-conclusions}.
%\end{enumerate}
\end{comment}
To address these questions, and as our first contribution, we conduct
a qualitative study, using interviews, to understand why and how
people defer messages and what affects their decision (Section~\ref{sec-definition}). As
our second contribution, we then present a detailed analysis of email
deferral by performing a large-scale log study of email interactions
of thousands of users (Section~\ref{sec-characterization}), that largely supports the trends
observed in our qualitative interviews. Inspired by our qualitative
and quantitative learnings, we define the task of predicting email deferral
from user interaction logs and describe a model for identifying
deferred messages using a large number of features (Section~\ref{sec-predictions}). Next
in Section~\ref{sec-discussions}, we discuss the implications of our study on designing
email clients. Finally, we provide an overview of the most related
work in Section~\ref{sec-relatedwork}, before we conclude the paper in Section~\ref{sec-conclusions}.

\section{Defining Email Deferral}
\label{sec-definition}
\vspace{-0.6mm}
%Siu et al \cite{Siu:2006}, through a day long in-situ shadowing study, observed that users explicitly flag emails for later action (e.g. messages left in the inbox marked  ``Unread'', message windows that were left opened on the desktop, or half-written replies that were visible on the desktop or saved in the  ``Draft'' folder).
%The shadowing study highlighted the prominent role of deferral as a frequently used email handling and task management technique.  %Circumstances often necessitate the need to delay email handling, leaving messages in a half read state or replies left partially composed.  

Siu et al \cite{Siu:2006} used a day long in-situ shadowing study, and observed that users explicitly mark emails for later action (e.g. mark messages as  ``Unread'', leave message windows opened on the desktop, etc.). 
%The shadowing study highlighted the prominent role of deferral as a frequently used email handling and task management technique.  
However previous literature has little information on how often users postpone required actions within their inbox during the day, what are the motivations behind this behavior and what strategies are used for revisiting these deferred messages, choosing instead to focus on other aspects of use such as triage or task management. 
Furthermore, there is no consensus on a formal definition for deferral, and little understanding on how it can be inferred from the user actions. 
To address these gaps, we conducted a qualitative study through a series of in-person interviews. The interview data helped us understand how deferral is perceived by enterprise workers, highlighted the motivations behind deferral and the strategies that are employed to facilitate getting back to deferred messages. %While deferral seemed to be very common among our interview participants and many factors were identified as key contributors for email deferrals, we conducted a follow up survey to validate these findings with a larger audience.

\vspace{-0.6mm}
\subsection{Interview} 
%\mss{
%demographics, questions 
%\paragraph*{Summary} 
%findings, quotes
%}
%In order to understand why and how deferral occurs for enterprise workers, how it is handled and whether it is possible to infer deferral behavior from users actions we conducted a series of semi-structured interviews at a large US-based technology company. 
We recruited 15 participants (4 female), all above 18, with a very diverse set of roles within Microsoft, ranging from product managers and researchers to software developers and interns. %\ms{do we have more information about demographics, such as age?}\bs{no, participants didn't need to specify their age. We just know everyone was above 18.} \footnote{\ms{we should acknowledge that our sample is still very biased towards young tech-savvy users.}}
Interviews were scheduled for 30 minutes in the participant's office where they had access to their mailbox during the interview.
All participants used Microsoft Outlook on a daily basis and some of them used other email clients as well. 
Part of the interview was conducted as a contextual inquiry, where the participants were asked to look through their mailbox and share some information about any emails that they had received on the day of interview and \textit{had to leave for later to deal with}.\footnote{ 
The participants were asked not to share any confidential information and use general terms to describe the content of those emails.}
We were mostly interested in learning about what those emails were about, what was the participant required to do in order to take care of that email, what was the relationship between the sender and the participants, whether there were multiple recipients on the emails, why did they decide to leave it for later to deal with, what would ``dealing with'' involve when the participant does go back to the email, how soon they are planning to do so, and finally whether they employed any strategies to facilitate this revisit.

For the remainder of the interview, the participants were asked a variety of questions, such as how many emails they receive per day, what kind of email management strategies they usually employ (e.g. whether they use many folders, have a clean inbox, have many unread emails, etc.), how often they leave emails that they have seen for later to deal with, and how they decide whether they should deal with an email right away or leave it for later. We also asked them to think about the actions they take on a deferred email at the time they go back to take care of it.

Interviews were all transcribed and analyzed using affinity diagramming. Data was initially clustered using open coding producing 16 clusters. Clusters were then analyzed using axial coding to identify overall themes in the data. 
We found that saturation for our qualitative data occurred approximately after 10 participants, no new clusters of information were identified. However, we continued to cluster interview data for the remaining participants, particularly attuned to data that might expand our clusters or add nuance to our analysis. These clusters resulted in three broad themes, namely: deferral is common; 
%deferral commonly involves response; 
factors that impact deferral; and revisiting deferred emails.
We describe more details about these themes in the next two subsections.
\vspace{-0.3mm}
\subsection{Main Findings}

%\paragraph{Deferral is common}
Our interviews confirmed that deferral is common,
and all our participants reported that they frequently
defer some emails.
%While our participants described a variety of email managements strategies and attitude towards handling their emails, a common theme among all of them was \emph{``not every email can be or should be handled right away"} [P10]. 
%They described many different scenarios where they decided to defer acting on email: \emph{``at the first moment I don't know what to answer or *if* to answer''} [P10, P11]; 
\emph{``I usually have many meetings during the day and so I just quickly look at the emails to see if it needs anything from my side, then leave it for later. This happens for half of my emails I would say.''} [P4]; \emph{``several times a day''} [P2, P11]; \emph{``at least 1 or 2 emails everyday''} [P9];

%\bs{ADD: participants who believed deferral is not common but were referring to long term deferrals.}
%While we had a group of participants that mentioned they usually manage to be on top of their emails and another group that felt it's very challenging to manage a never ending flow of emails everyday
% While those users who manage to be on top of their emails usually do handle most of their emails by the end of the day, deferral does happen commonly and not everything can be handled right away.

%\paragraph{Factors impacting the decision to defer}
%As mentioned, the first half of our interview involved participants going through the emails they received on the day of the interview and answer different questions about those messages. This approach helped us gather more grounded responses without the need to access the participants mailboxes to ensure high ecological validity.

Participants discussed a wide range of characteristics regarding the emails they deferred as well as other factors that impact their decision to defer emails during triage sessions. 
These factors can be grouped into 5 main classes, ordered by prominence as: (1) How much time or effort does handling this email require? (2) Who is the sender? (3) How many recipients are on the thread? (4) What is the user's workload and context? and (5) What is the urgency of the email message?. %\ms{Is this ordered based on the frequency of responses from participants? If not, let's sort the list and the corresponding content later in this section.}\bs{it makes sense to swap Urgency and Workload. But the rest are in order of questions in the interview as well as frequency of responses.}

\paragraph*{Time \& effort} The amount of time or effort needed for handling an email was identified as the primary factor influencing the decision to defer an email to a later time. Different characteristics of the email were used to estimate the required time or effort at the time an email was first seen by the participant: Do I know the answer? [P7, P11, P13] Does it require any task to be done? [P4, P12] What is the complexity level involved? [P11] Does it require context switching? [P4, P11, P13] Can I handle it independently? [P1, P11]

\paragraph*{Sender} The sender of an email and the relationship between the sender and the recipient was another factor influencing deferral. Participants mentioned different notions of `importance' for a sender: based on the projects they are affiliated with [P9, P11]; based on their organizational rank [P2, P11, P12] or based on their proximity [P11, P12]. %\ms{Earlier in the paper we suggest that the sender is not very important. Perhaps we should rephrase that part a bit.}\bs{that's right! while workload is more important than the sender the main finding was that the amount of time/effort required for handling an email was by far the main factor. So we had many quotes saying that if handling an email is easy they just do it, regardless of who sent the email.}
Time zone of the sender could also lead a recipient to prioritize handling the email in two different ways: \emph{``If I receive emails that I know it's like 3-4 pm their time, I try to answer those first because then they can move on with their day before work hours are over. %Emailing with different time zones needs more planning!''
} [P10]; \emph{``And the ones that I'm the only recipient then I usually try to answer as fast as possible. unless I know that the other person is sleeping right now. 
%Then there is no rush.''
} [P6] 

Finally ``Intended Responsiveness Image'' emerged as another factor that some of the participants considered while deciding how quickly they need to respond to an email that they just saw. As identified by \citet{Tyler:2003}, users often maintain and cultivate a responsiveness image for projecting expectations about their email response. \emph{``There are times that the relationship between me as the recipient and the sender is such that I don't want to respond right away. Sometimes it's because the sender is not very important and sometimes it's because I don't want to come across as too available"!} [P2]
%\emph{``Right now I try to reply right away to everybody because I'm new to the team and I'm trying to make an impression!''} [P11]
\paragraph*{Recipients} Having multiple recipients on the thread commonly led to a delay in handling the email for two different reasons: If there is an overlap in knowledge and ability in handling the email a recipient might choose not to act knowing other recipients can also take care of it [P6, P9]. Different participants mentioned the need to be able to identify whether or not \textit{they} are expected to handle the email: Can I immediately tell if \textit{I} need to take care of it? [P7, P13]; Am I explicitly mentioned? [P12, P13]; Am I on the To line or the CC line? [P7, P12].
Another related case was active email threads with multiple recipients on them: \emph{``there are many cases with those threads where multiple people have already responded and I should read it for a while to see who said what.''} [P9]; %\emph{``so most of our email threads by the time they reach me they're about thirteen emails long and I'll have to stay informed.' } [P12];
%\emph{``The other email I haven't had a chance to read throughly already has so many replies to the thread so I need to spend some time reading it later.''} [P4]. % \ms{Shouldn't this be discussed under "effort"? Although, I can see why we have it here too.} \bs{yes it belongs to both themese and I wasn't sure where to put it.}
\paragraph*{Workload} The context and workload of the user was identified as an external factor impacting the deferral decisions at the time of first visit to an email. The context of a user involved the workload (e.g. number of pending tasks, number of unhandled emails), current task at hand (e.g. coding, attending a meeting), their whereabout (e.g. at office, at home, in transit) and access to resources (e.g. having access to a large screen, having their planner handy, etc).  %\ms{Again, can we back up with some quotes from the interview? We have good content on sender and recipient but it's pretty shallow for the rest. Is it consistent with the responses we got from participants?}\bs{added}
\emph{``I use conditional formatting to know which emails I am on the To line vs CC line and if the email mentions me.  So the issue is not identifying those emails, the main problem is when I don't have time to deal with them. e.g. running into a meeting. So I have to flag them and remember to go back to my flagged emails later.} ''[P4]; %\emph{``If I see an email but I'm on my way to a meeting then I'll definitely leave it for later.}'' [P9]; % \emph{``The other email I received when I was in a meeting and it needed a response. And I needed to connect with one of my colleagues to be able to respond. So I left it for later.''} [P11]; \emph{``generally it's because I have something else on my plate right now that should be finished first before I have time for handling the email.''} [P6]; 
\emph{``generally it's because I have something else on my plate right now that should be finished first before I have time for handling the email.''} [P13]. 
%\emph{``Sometimes these emails are important but it requires me to contact somebody or do some extra work. So let me see what the other things I need to do are first and do them before I go back.''} [P2].
%		a. Am I in a meeting / walking to a meeting?
%		b. Am I in transit / checking the email on my mobile device vs Sitting at my desk?
%		c. What current task I'm engaged in?
%			i. Does handling the email require context switching?
%		d. Workload
%			i. How many pending tasks I have for today?
%			ii. How many other emails I need to deal with?
%			iii. How many meetings I should attend?
		
% Am I blocking someone?

\paragraph*{Urgency} we learned that the urgency of an email was perceived in at least two different ways: urgency for the recipient of the email versus that for the sender. For both notions the participants would evaluate the urgency of an email based on (a) whether a deadline is specified in the email, and (b) if there is a deadline for the task / project with which this email is affiliated. %\ms{can we back up with some quotes from the interview?}\bs{added}
%\emph{`` if there is a faculty meeting that is coming up in a week and I need to prepare the agenda, topics to discuss, etc; And it certainly needs a bit of a quicker reaction. So I have to answer within a day. And it's also expected somehow for these critical things to get a reply fast.''} [P10] 
\emph{``[How soon I revisit deferred emails] would depend on their urgency. I keep that in mind as scrolling past them and thinking they won't need that just yet. %But normally I do go back to them within a day or two.
}''[P7]; 
\emph{``I would rather read all of them and then start working on them than start working on the ones that I have already read; because something that could be five emails down could have a one hour deadline, whereas something that just came in may have a one day deadline.''} [P12]. %``There are many emails that are like: hey this is what I did, can you give it a look and let me know how it looks? those are not so much of a priority for me but I make it a priority for the team because that person's usually doing the presentation somewhere or presenting to one of our partners and they are on a tight schedule.'' [P12]}

\subsection{Characterizing Revisits}
Based on the qualitative data that we collected, there are three main aspects to revisiting deferred emails: \textit{how}, \textit{when}, and \textit{what}. The first aspect (how) is about strategies employed by users for deferral. The second aspect (when) is related to factors that influence the time of revisit. Finally, the third aspect (what) is regarding the actions taken by users when handling their revisited deferred emails.
%While the former involves deferral strategies, the latter has a lot to do with task management and scheduling. 

%Facilitate happens at the read time
%When are you going to revisit (e.g. reply)
%What actions happen during your revisit (e.g. move)

\paragraph{Deferral strategies (How)}
Deferral Strategies seem to be affected by the users' email management attitudes and behaviors. Previous studies have demonstrated the different strategies employed by email users to manage messages. \citet{Whittaker} classified users into frequent filers, spring cleaners, and no-filers, which was later extended by  other classifications (\cite{Gwizdka,Kalman}).
Similarly in our study we identified a connection between the choice of deferral strategy and the user's general inbox management attitude. For example, \textit{zero-inbox}\footnote{\url{https://www.goodreads.com/book/show/8660916-inbox-zero}} users and filers who frequently move emails out of their inbox do use their inbox as a TODO list and leave their pending deferred emails in the inbox as a reminder to go back to them. \textit{Zero-unread} users who actively try to minimize the number of unread emails in their inbox do use \textit{MarkAsUnread} as a strategy to facilitate finding the pending emails to go back to them. Although \textit{flags} were used by a variety of users types, it was perceived differently by pilers, zero-inbox or zero-unread users. While pilers seemed to use flag as a way to remember about pending items that are usually mixed with many other unread emails in the inbox, zero-inbox or zero-unread users usually apply flags to indicate an important or urgent email.
Some participants used a mix of email functionalities to mark pending emails that require a response while others used external tools such as OneNote, Planner, Calendar, etc. to plan for and schedule their pending tasks that were corresponding to their pending emails. 
%For example, 
\emph{``So my general strategy [during Triage] is I flag emails that need a response. If a task needs to be done I add it to my schedule as a block in my calendar.''} [P4].

\paragraph{Scheduling (When)}

%A related dimension to how users go back to deferred emails was the temporal aspect of deferred emails revisiting. 
Participants often took a variety of factors into account to determine the time to go back to a deferred emails. Two main factors were common across all participants: (1) estimating the amount of time, prerequisites and resources required to handle that email, %(2) estimating the required prerequisites and resources to be able to handle the email, 
and (2) finding the best available time slot in their schedule to handle that email.
Focusing on the prerequisites of handling an email, participants talked about a variety of such requirements including whether or not they need to consult with a colleague [P10, P11], what sub-tasks handling this email involves [P9, P12] and where they should be to handle this email [P8, P13]. %\ms{sample quotes? also, we haven't really covered time here.} %\bs{added some. can add more later.} 
%\emph{``But if the Vice President emails me for faculty hiring or decision to fund some project, I need to think about it. Or I need to consult with my associate deans. So it can create a loop. Untill I can actually answer.''} [P10]; \emph{``Sometimes the content of the email alone is not enough to take care of the task it requires me to do. I use other tools like OneNote to write down things I need to do in order to take care of this email.''} [P9]. %\emph{``normally if I'm at my desk like if I'm in the office but I'll just type up a note but it's hard to do that while reading an email on the phone; so I left it there figured I'd re-read it when I got into the office and make a note that.''[P13]}

Even after the user has determined the time and the prerequisites to handle an email or the tasks associated with it, finding the right time to do it turned out to be a different process that involves optimizing two different variables: (a) minimizing context switch, and (b) minimizing the number of time slots to be assigned to handling a single email.
\emph{``a lot of people find that most of their energy is spent switching gears and so if you want to minimize the amount of gear switching then every time you flag something for later you should indicate what kind of later it is. 
%For example, there are forms that I can fill out when I'm at my computer; energy bill I should do when I'm at home, etc. 
''} [P13]. 
\emph{``So when I have 30 minutes I want to devote it to something longer. [...] and then for easy tasks I'm gonna do it when I have five minutes and I won't do it when I have 30 minutes.''} [P12].
%\ms{Should we again cite and briefly compare the previous work on re-visits in this section?}

% \begin{enumerate}
% \item How
% \begin{itemize}
% \item Affected by the User's Type
% \begin{itemize}
% \item Email functionalities (e.g. Flag, MarkAsUnread, TODO folder, Inbox as a TODO list)
% \item External tools (e.g. OneNote, Planner, Calendar, ...)
% \end{itemize}
% \end{itemize}
% \item When
% \begin{itemize}
% \item Estimating the time needed to handle the email
% \item Estimating the required pre-requisits / resources
% \begin{itemize}
% \item Consult with a colleague
% \item Sub-tasks to be done
% \item Where should I be to handle it?
% \end{itemize}
% \item Where in my day does it fit?
% \begin{itemize}
% \item Minimize context switching
% \item Minimize the number of available time slots to assign (e.g. handing 3 5-min tasks vs a 15-min task)
% \end{itemize}
% \end{itemize}
% \end{enumerate}

\paragraph*{Handling deferred emails (What)}
As mentioned earlier, one of the interview questions was about the actions participants take when they go back to a deferred email. We collected responses for both the specific emails that were deferred on the day of the interview as well as the more general discussions regarding the past deferral cases.
The \textit{need to respond} was identified as the \textit{most common action} users take when they revisit deferred emails: \emph{``[the main reason for deferral] is just that I wanna think about this a little bit, so I let it roll at the back of my mind to see how I should respond to this.''} [P2]; 
\emph{``
%At the time that I go back and then actually open the email and more or less I know what I want to write. so 
basically going back to emails almost always involves responding.''} [P6]; 
\emph{
%``Yes and when I go back to it I still need to reply. 
yeah if it's unread it's because I need to do something with it. And when I go back to it I pick up the conversation thread and reply.''} [P5]. 

While all participants talked about ``response'' as the most common action they take for their pending emails, this response is not always a "Reply" action on the deferred email; %Replying to the same email was still the most common scenario, but 
composing a new message or a face-to-face or phone conversation were also mentioned as ways to follow up and resolve a deferred case. 
\emph{``
%When I get overloaded and I'm trying to take care of other things and I'm behind on email, 
it's definitely a case that I choose not to respond [to a deferred email] if it becomes more urgent for that person then they email me back or come to my office and get things resolved more efficiently.''} [P7];
\emph{``if it's something that says hey we need to organize something I don't need to respond to that email. I just set it up and then [...] 
%when I'm certain that I've set it up correctly 
I start a new thread saying that.''} [P12].

We also recorded a few scenarios where the user needed to defer an email to be able to read it for a extended period of time. The most common scenario with \textit{long read} involved active threads of emails with multiple recipients on them. Some of these cases would still result in a reply to the thread, while sometimes the user just needed to stay informed.  
% Long read came up only a few times and in most cases involved active threads of emails where multiple recipients are involved and multiple replies are already done before a recipient sees an email for the first time.
%Other cases with long read was included a task management scenario where a user needed to integrate a document that is shared via email to his notes, or when a user needed to go through a code that was sent out by email to see if he needs to use it for his project. Both of these scenarios, while recognized, were not the case for the majority of the participants.

%\bs{Summarize and list all these actions: Reply --> the dominant case followed by Forward (case of deferral by delegation) and Long Read (case of active threads and task management}

%\mss{
%\subsection{Survey} 
%demographics, questions 
%\paragraph*{Summary} 
%findings
%}

\subsection{Summary}
%\mss{summary of findings and practical considerations (e.g. ignoring reads for the next section)}
%\bs{I suggest we formally define our Research Questions in Section 3 and use the Qualitative data collected to answer them. Then here we can revisit them and draw practical conclusions that can inform our next steps; that is (1) quantitative validation and characterization of deferral and (2) inference and prediction task.}

%Understanding how users defer messages and the conditions surrounding its use is crucial because it represents an entry point into the user's future task list.  Obtaining data on this mode of use therefore was the focus in the second half of our research, which tried to gather quantitative measures of email interactions in hopes of capturing a long term picture of deferral in action. 

%\textbf{[RQ1]: How is deferral Perceived by different users?\\}
In our qualitative study, we directly investigated each of the research questions outlined early in this paper.
%\sd{It seems odd to introduce new methods/data in the discussion, although this paragraph may be equally out of place at the beginning of the mail findings section.}\bs{I see your point. Yet I'm not sure where else we can talk about RQ1 after the entire process is described?} 
Our interview questions were designed to minimize priming the participants with our definition or specifications of deferral. To this end, we referred to the concept of deferral as ``emails that are seen but were left for {later} to {be dealt with}.'', while the terms ``later'', ``dealt with'' and the reasons behind this behavior was open to interpretation by the participant.
Participants differed in their perception of what ``later'' means for a deferral case. The majority of participants considered a message left for a day as deferral. Short term deferrals were also common among participants; especially when they were specifically asked about them. We also learned that in almost all cases of deferral, the user intention is to return to an email and \textit{complete the
task}. That often involved taking a strong action like replying to the email. Therefore, we define an email as deferred after being read for the first time if:

\begin{quote}
the user deliberately postpones completing the task related to it to later (RQ1).
\end{quote}

Our interviews confirmed that deferral is  common and was reported by
all fifteen participants in our study (RQ2). We learned that users defer emails mostly due to 
lack of sufficient time for handling emails and related tasks that may involve synthesis and 
collecting resources (RQ3). Indeed, deferral was mentioned as an effective time and task management technique for reducing context switching. Our participants pointed out several factors
that influence their decision about deferring an email, including the time and effort, sender
importance, number of recipients and previous emails on the thread,
their workload at the time of reading the email for the first time, and urgency of the email (RQ4). We discovered that users
apply various strategies to stay on top of their deferred emails and tasks (RQ5). Marking emails as unread after reading, came up as of the most common strategies, which confirms previous related findings \cite{Siu:2006}. Creating dedicated folders, or transferring tasks to external tools, such as note taking tools, were preferred by some of the participants, although they were not as common.

In the next section, we revisit the same questions through a different lens, by characterizing deferrals in the user action logs of a commercial email provider.

%\paragraph*{Defining Deferral:\\}
%Pending emails that are left for later and require action. While revisit times and reasons behind deferral varies from user to user deferral is common, challenging and requires support from the system side.

\section{Characterizing Email Deferral}
\label{sec-characterization}
%\textbf{[RQ2]: How common is deferral?\\}
%\textbf{[RQ3]: High level Motivations behind Deferral?}
% \begin{enumerate}
% \item Lack of sufficient time for handling emails;
% \item Effective Time and Task Management
% \item Minimizing Context Switch
% \item The need to synthesis, tasks to be done and resources to gather;
% \end{enumerate}
Our detailed in-person interviews highlighted several interesting aspects of why and what emails are deferred. However, obviously our findings, based on 15 interviewees, should not be extrapolated to all users.
%\footnote{The interviewees who agreed to talk to us were clearly biased towards power users.} 
Therefore, as in the previous section, we study the research questions described in introduction, but this time quantitatively based on user action logs.\footnote{Please note that (RQ3) cannot be studied using action logs and hence is not explored here.}

%\subsection{Data} 
We analyzed a sample of the anonymized email logs from users of a major commercial Web email client (Outlook) over a two weeks period from May 6, 2018 to May 19, 2018. The email Web client can be used on both desktop and mobile with multiple browsers. 
Our sample consists of forty thousand active users who performed about 3 million actions during the two weeks period. 
%The typical configuration has a folder list on the left and a search box is on the top left, the message list in the middle, and the message reading pane on the right.
Our sample included emails from enterprise users only. The logs do not contain the text of the email message, email headers or email search queries. The email log contains actions performed against messages with timestamps and other metadata. We limit our analysis to \textit{active} users by excluding any user that interacts with less than 1\% of received email in the two week period.

\subsection{Deferral in Email Logs (RQ1-2)}
In the previous section, and based on our qualitative study, we defined an email as deferred when "the user deliberately postpones completing the task related to it to later" (RQ1). 
To quantify how often deferral can be observed in email action logs (RQ2), we first need to adopt
a definition that can be measured based on traces of recorded user actions. To achieve this we have to identify what signals can be considered as proxy for \textit{task completion} and \textit{later}. To address the former, we decided to focus entirely on three \textit{strong actions}: Reply, ReplyAll, and Forward, which were frequently mentioned by our interviewees as the main unhandled
tasks related to their deferred emails. To set the threshold for what qualifies as \textrm{later}, we were again inspired by the responses from our interviewees who suggested that they would get back to their deferred emails later in the day. 

Therefore, we limit our focus to emails on which a strong action is observed only in \textit{sessions} after the initial read session. While session boundaries are well studied and relatively well understood in some related areas such as Web search \cite{Jones:2008:BST:1458082.1458176}, there is very little work on how they should be defined for email triage. Following the work on session segmentation in Web search, We segmented our logs into session based on a threshold on the time of inactivity between actions. Following~\citet{Narang}, we use 10 minutes of inactivity as our threshold.\footnote{It is worth noting that the focus of \citet{Narang} paper was on email search. We explored a few other thresholds in our preliminary experiments and observed little difference. Hence, we left further investigations for future work.} 
Based on this definition, and using our sample of email logs, we found that 16\% of active users defer at least one message per day (excluding weekends), 3\% of all messages get deferred, and that deferral happens in at least 12\% of triage sessions. Overall, deferral is indeed common based on both qualitative interview data and quantitative email logs (RQ2). 

\vspace{-0.3mm}
\subsection{Factors Impacting Deferral (RQ4)}
We learned from our interviews the decision to defer an email can be impacted by several factors such as the properties of the email (e.g. sender, receiver) and the workload of the user. We investigate these factors using log analysis here.

\paragraph*{Properties of deferred emails}
%Next we looked at how the messages characteristics are impacting deferral. 
Using our qualitative data we identified four main characteristics of messages that usually lead to deferral: (1) {time \& effort} (2) {urgency} (3) {sender}, and (4) {recipients}.
%number of recipients on the email. 
Among these, the first two are best inferred from email content, such as subject and body, to which we had no direct access in our logs.
%Since \textit{effort} and \textit{urgency} are usually inferred from the content / subject of the email and we do not have access to any textual information in our dataset, we used existing classifiers that 
We could however run text classifiers blindly on the body of emails to detect whether an email is requesting any action (isActionRequest) from the recipient(s), and whether the email content is requesting a reply (isReplyRequest) \cite{Yang:2017}. We use these two signals as a proxy for ``the amount of work required'' to handle an email. We did not however have any indication of \textit{urgency} of messages available in our dataset.
We also used algorithmically computed metadata about the sender (i.e. Human vs. machine-generated (\textit{isHuman}), whether the sender is important based on historical interaction features (\textit{isImportantSender}) and whether the sender has sent any emails to the recipient(s) in the past (\textit{isKnownSender}). We also used the number of recipients (\#recipients). 
%We use these features to test H4 and H5.
%\aha{how is isimportant defined}

Table ~\ref{tbl-properties} presents the distribution of these features given the type of message. %We included a few other available features for completeness. 
It can be observed that Non-Deferred messages tend to have double number of recipients on average (\textit{recipients}), and contain noticeably fewer number of action or reply requests (proxy for effort) than Deferred messages. Furthermore, a Non-Deferred message is less likely to be from a human or important sender and more likely to be from a known sender. % the difference is not *significant* (\textit{sender}).
%These results supported H3 and H4, while H5 is not confirmed. \bs{statistical test needed to complete this.}

\begin{table*}[htb]
\centering
\caption{Comparing the properties of deferred and non-deferred emails based on the meta-data
and contents of messages. % averaged over each group of emails. %Statistical significant differences ($p<0.05$) are distinguished by \underline{underline}.
}\label{tbl-properties}
\vspace{-10px}
\begin{tabular}{ l | l l l l l l }
  \hline
  Type & \#recipients & \#actionRequests & \#replyRequest & isHuman & isKnownSender & isImportantSender \\
  \hline
  Deferred		 & {3.899} & 0.075 & 0.200 & 0.849 & 0.604 & {0.469} \\
  Non-Deferred 	 & 7.010 & 0.034 & 0.100 & 0.744 & 0.723 & 0.403 \\
  \hline
\end{tabular}
\end{table*}

\begin{table*}[h]
\centering
\caption{Probability of observing a given action on an email (deferred vs. non-deferred), and first read vs. revisit sessions.  %Statistical significant differences ($p<0.05$) are distinguished by \underline{underline}.
}\label{tbl-read-revisit}
\vspace{-10px}
\begin{tabular}{ l | l l l l l l l l}
  \hline
  Type & Delete & Flag & FlagComplete & LinkClicked & MarkAsUnread & Move & OpenedAnAttachment \\
  \hline
  Deferred-Read		 & 0.004 & 0.021 & 0.001 & 0.017 & 0.038 & 0.015 & 0.139 \\
  Deferred-Revisit 	 & 0.054 & 0.005 & 0.011 & 0.014 & 0.008 & 0.086 & 0.096 \\
  Non-Deferred-Read 	 & 0.121 & 0.007 & 0.003 & 0.034 & 0.008 & 0.060 & 0.087 \\
  Non-Deferred-Revisit 	 & 0.054 & 0.003 & 0.003 & 0.008 & 0.004 & 0.0300 & 0.027 \\
  \hline
\end{tabular}
\end{table*}

\paragraph*{The Impact of workload on deferral}
%email workload, and \#meetings. Figure~\ref{fig-workload}...
The workload of a user can be estimated using a large set of factors including the projects they are involved with, the meetings they have to attend or prepare for, the number of emails they receive and hence need to triage, etc. Since we do not have access to a comprehensive context of the user, we use two factors to estimate the workload of a user at the time she is attending to her mailbox: (1) number of unhandled messages that she received in her inbox since the last time she visited her mailbox and (2) number of meetings or appointments the user has on her calendar during the time of receiving emails. %\bs{check with CJ / Use what Chris wrote for Section 5}
%To elaborate, the meeting workload reflect the user's calendar at the time of the initial read of the message. 

Such features may be helpful because what users are doing may affect how they interact with their email. For example, as mentioned by many of our participants, they tend to defer more emails if they are in a meeting when they first read the email. For every message, we look at the hour and day of the initial read of that message and include counts of the total number of meetings at that time using the user's calendar data.
%\aha{you mean the total number of meetings for the rest of the day starting this time?}
%with attendees, the number of meetings with attachments, and number of meetings for which the user is an organizer, the number of meetings total, the number of appointments total, and the number of markings that show the user as free, busy, tentative, or out of office.

\begin{figure}[tp]
\includegraphics[width=2.5in, viewport = 60 70 720 530,clip]{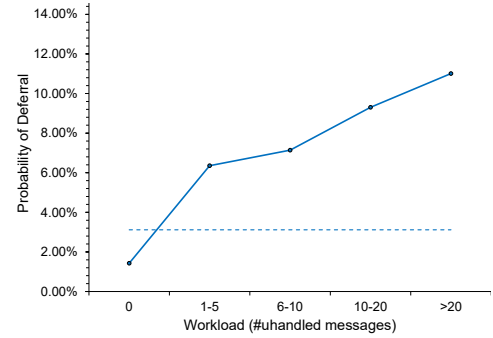}
\caption{The probability of deferring an email based on the workload of the user measured by %(1) 
the \#unhandeled emails % - top, and (2) the number of meetings on the calendar - bottom. Dashed line on the tip figure represents the probability of deferral for all workloads
}\label{fig-workload}
\end{figure}

Figure \ref{fig-workload} illustrates the relationship between the workload of the user in terms of number of unhandled emails and the probability of deferring a message. %In both figures t
The horizontal line indicates the baseline deferral probability regardless of the workload of the user. The trends are consistent with our qualitative findings and confirm that users are more likely to defer a message when dealing with higher workload. 
Similar trends -- omitted here for brevity -- were observed when number of meetings is considered as proxy for workload. For instance probability of deferral increases from about 3\% to 4\% as the number of meetings on user's calendar almost monotonically grows from 1 to 5.

\vspace{-0.5mm}
\vspace{-0.3mm}
%\ms{Deferred vs. replied vs. revisited 
%Plots and tables based on log analysis, connection to previous sections. We'll probably have different paragraph here each focusing on a particular analysis}
% workload meeting
%1	3.14%
%2	3.28%
%3	3.23%
%4	3.87%
%overall 3.17%
\vspace{-1mm}
\subsection{Deferral strategies (RQ5)}
%\% of emails classified as deferred vs. not deferred using our definition
%Our sample contains about about 1.3 million distinct messages among them 3\% are deferred using our definition.
Our qualitative findings showed users employed a variety of strategies to manage their deferred emails. While inferring high-level management strategies from actions logs is a challenging task, actions such as Flag and MarkAsUnread were commonly mentioned as deferral strategies (\textit{deferralStrategy}), whereas Move was mostly used for archiving a non-deferred message or filing a deferred message when it was no longer pending (\textit{filingStrategy}). We used our action logs to compare the probability of observing these actions %-- as proxy for deferral strategies -- 
given the message type (i.e. Deferred vs. Non-Deferred).

Table \ref{tbl-actions} includes these conditional probabilities for these actions along with other actions for completeness. We compute standard error confidence intervals for all probabilities in Tables 2-3 via bootstrapping, and find that all differences are statistically significant with $p<0.05$. %We also show the odd ratio of observing theses actions for deferred vs. non-deferred messages in the top portion of Figure~\ref{fig-actionsodds1}.
Looking at \textit{Flag} and \textit{MarkAsUnread} columns we see that they are indeed more likely to be observed for Deferred messages than Non-Deferred ones, which partially validates RQ5. However, \textit{Move} seems to be used on Deferred messages more frequently which is in contrast to our qualitative findings -- a discrepancy we return to later in this section. 

\begin{table*}[htb]
\centering
\caption{Probability of observing a given action on an email given its type (deferred vs. non-deferred) . 
%and surveys as common actions related to deferral. %Statistical significant differences ($p<0.05$) are distinguished by \underline{underline}.
}\label{tbl-actions}
\vspace{-10px}
\begin{tabular}{ l | l l l l l l l l}
  \hline
  Type  & Delete & Flag & FlagComplete & LinkClicked & MarkAsUnread & Move & OpenedAnAttachment \\
  \hline
  Deferred		 & 0.108 & 0.036 & 0.021 & 0.032 & 0.053 & 0.161 & 0.239 \\
  Non-Deferred 	 & 0.171 & 0.009 & 0.006 & 0.039 & 0.011 & 0.090 & 0.103 \\
%  ImmediateSA 	 & 0.134 & 0.012 & 0.008 & 0.016 & 0.013 & 0.135 & 0.114 \\
  \hline
\end{tabular}
\end{table*}

In order to further investigate RQ5, we look at the distribution of actions in read and revisit sessions separately. The motivation behind this analysis is that the intent of the user at the time of first read is indeed different from the time she goes back to a message at a later time. While the intent to defer a message takes place at the time of first read, the intent to take care of a pending message takes place at the time of revisit and can similarly impact the distribution of user's actions.
This hypothesis is supported by our qualitative data as users seemed to exhibit different behavior at the time of the first reading of a message (e.g. \textit{deferralStrategy}) and the time they revisit that message (e.g. \textit{filingStrategy}). 
%For example, looking at deferral strategies, they were meant to facilitate the revisit to a deferred message and hence the user would apply it at the time of first read. 
%Our interview data also included actions that users take at the time of revisiting messages.
In order to investigate the differences between read and revisit activities, 
we split our distributions of actions into first read and first revisit sessions.  

Table \ref{tbl-read-revisit} summarizes the probability of observing different actions in read and revisit sessions and contrast them between Deferred and Non-Deferred messages. %Additionally, we show the odds ratio of observing different action in the read and revisit sessions of deferred messages in Figure~\ref{fig-actionsodds2}.  
While it is still evident that \textit{Flag} and \textit{MarkAsUnread} are more likely to be seen for Deferred messages, we can see that they are indeed used as deferral strategies at the time of reading these messages. Additionally, looking at the read sessions of Deferred and Non-Deferred messages, we see a significant difference in the likelihood of observing these two actions which is consistent with our qualitative findings for Q5 (\textit{deferralStrategy}).

Another interesting observation is the distribution of the \textit{Move} action. At the time of reading a message with the intention of deferral, observing the \textit{Move} action is the least likely, whereas users are more likely to move a message at the read time than the revisit time if they are not intending to defer acting on it. Furthermore, \textit{Move} at the time of revisit for a deferred message seems to be an indicator of ``being done acting on that message and hence the message is no longer pending''. This behavior was also observed among our participants and more specifically for the Filers who tend to move emails out of their Inbox as soon as they are done handling them (\textit{filingStrategy}), and explains the discrepancy observed earlier in this section.

\begin{comment}
\begin{figure}[tp]
\includegraphics[width=3in, viewport = 60 70 720 520,clip]{OR-Def-NotDef}
%\includegraphics[width=3in, viewport = 60 70 720 520,clip]{OR-Def-NotDef-Read}
%\includegraphics[width=3in, viewport = 60 70 720 520,clip]{OR-Def-Read-Revisit}
%\includegraphics[width=3in, viewport = 60 70 720 520,clip]{OR-NotDef-Imm-NoAct}
\caption{The odds ration for the probability of an action given its type: Deferred vs. Not-Deferred.
}\label{fig-actionsodds1}
\end{figure}
\end{comment}

\begin{comment}
\begin{figure}[tp]
\includegraphics[width=3in, viewport = 60 70 720 520,clip]{OR-Def-Read-Revisit}
\caption{The odds ration for the probability of an action on a deferred message given the session: Read vs. Revisit.
}\label{fig-actionsodds2}
\end{figure}
\end{comment}
\vspace{-0.7mm}
\subsection{Deferral vs. Reply}
Given that all deferred emails in our dataset are associated with a strong action, it is important to verify how they are different from other replied-to emails, and confirm if there is anything unique about deferred emails. 
%\sd{I found this difficult to follow.  Is there a way to add something to Tables 1 and 2 to summarize this?}\bs{adding a table on message properties and actions distribution associated with Replied emails could elaborate on this comparison. However, I do not have the data to make such a table. Perhaps Ahmed could help?} 

Interestingly, we observed that replied-to emails are much more similar to Non-deferred emails in terms of their action probability distribution, but their metadata and other properties are closer to Deferred emails.
For instance, the probability of \textit{MarkAsUnread} is 0.013 for replied-to emails (compared to $0.011$ and $0.032$ respectively for Non-Deferred and Deferred as shown in Table \ref{tbl-actions}). In contrast, the average number of recipients for replied-to emails is $3.497$ (compared to $7.010$ and $3.899$ respectively for Non-Deferred Deferred emails, as presented in Table \ref{tbl-properties}).We observed consistent trends for other actions and properties and exclude them here for brevity.

%we can see that the characteristics of messages with strong actions are far more similar to deferred messages and not the non-deferred ones. 

One way to explain this is that message properties may be good indicators of whether the user will take a strong action (e.g. Reply) on the message or not (as was also investigated in Yang et al. \cite{Yang:2017}). However, they may not be a strong indicator of determining whether the strong action will be taken immediately or deferred for later.

\section{Deferral Prediction Problems}
\label{sec-predictions}

We now present our experiments for predicting deferral. The ability to accurately predict whether an email is deferred has the potential of significantly improving the email experience. For example, email clients could use such a model to remind users about emails that they have deferred or even forgotten about, reducing the amount of effort they need to spend to re-find emails and the chance of them missing to act on important emails.

Since we do not have access to ground truth labels, we focus on emails that  have been marked as unread or flagged (the strongest signal of deferral that we have) and predict whether the user will return to them or not. More specifically, we predict whether a message that has been marked as unread or flagged will receive a strong action (Reply, ReplyAll, Forward) in a session after the session containing the email's initial read. We choose to only predict revisits with strong actions (as opposed to all revisits) because our qualitative study shows that almost all deferred emails contain strong actions in the revisit.

Note that the problem we tackle here differs from Reply prediction, because of the element of time. We do not build a model to simply predict whether a user will reply to an email \emph{at any time}.  Instead, we build a model to predict whether a user will return to an email and and take a strong action on it \emph{at a later time}. Our problem formulation allows us to use signals from the email's initial read session in order to help in the prediction. For example, a flag on the email could be a strong signal that the email will be returned to. 

\paragraph*{Labels \& features}
We first filter the sample of  email logs we use in our quantitative analysis to include only those emails marked as unread and flagged, resulting in a dataset containing 10,551 emails. We label messages as positive when they contain strong actions in a session after the initial read session. For our first experiment, we remove messages that contain strong actions in the initial read session because we do not consider these emails as deferred. We will return to consider these excluded examples in the following experiment in Section \ref{deferral}. 

We use a large number of features spanning different categories (see Table~\ref{tab:features}). \textbf{Action Features} include counts of user actions (e.g. Reply, Move, etc.) that occurred during the initial read session for that message. We also include \textbf{Message Features} like the length of the email body (in number of words), number of recipients and several features characterizing the sender of the message. \textbf{User Features} represent general user characteristics like the size of their mailbox (as a proxy for the amount of email messages they need to handle). We use four buckets to represent the number of emails they received in total in the entire dataset (<10, 10-24, 25-99, $\geq$ 100). We also use a feature to describe their  triage behavior (whether they are a Zero Inboxer, a Zero Unreader, or a Piler \cite{Whittaker}). We have shown earlier that the user workload is an important factor when she decides whether to defer a message or handle it immediately. Hence, we also included \textbf{Workload Features} to represent the user's workload in terms of messages they need to handle and meetings they have. Finally since the time of first seeing the message may have an impact on how a user handles it, we include several \textbf{Temporal Features} characterizing the time of the first read of the message (e.g. hour of the day, day of the week, etc.). A full list of features and their description is listed in Table~\ref{tab:features}.

\begin{table}
\caption{Features used for deferral prediction.}
\vspace{-1em}
\begin{small}
\begin{tabular}{ l | l }
  \hline
  \multicolumn{1}{c}{\textbf{Name}} & \multicolumn{1}{c}{\textbf{Description}}\\
  \hline
  \rowcolor[gray]{.9}\multicolumn{2}{l}{Action Features} \\
  \hline
  NumRespone & Num. of responses (Reply, ReplyAll, Forward)  \\
  NumFlag & Num. of flag actions (Flag, FlagComplete) \\
  NumMarkUnRead & Num. of MarkAsUnread actions \\
  NumOpenAtt & Num. of OpenAnAttachment actions \\
  NumLinkClK & Num. of LinkClicked actions \\
  NumMove & Num. of Move actions \\
  NumDelete & Num. of Delete actions \\
  NumSearch & Num. of times retrieved in search \\
  \hline
  \rowcolor[gray]{.9}\multicolumn{2}{l}{Message Features} \\
  \hline
  UniqueBodyLength & length of the body of the email \\
  isBulkMessage & Is email sent to a dist. list? \\
  isInThread & Is email part of a thread? \\
  numRecipients & number of recipients in the email \\
  isHuman & Is sender a human (vs. machine-generated)? \\
  isSenderFromSameOrg & Is sender from the same org? \\
  isKnownSender & Is sender known (sent emails before)? \\
  isImportantSender & Is the sender important (previous interactions) \\
  
  \hline
  \rowcolor[gray]{.9}\multicolumn{2}{l}{User Features} \\
  \hline
  MailboxSize & Num. of receieived emails in the dataset \\
  ManagementStyle & Piler, ZeroInbox or ZeroUnread\cite{Whittaker} \\
  \hline
  \rowcolor[gray]{.9}\multicolumn{2}{l}{Workload Features} \\
  \hline
  NumMessages & num. of unhandled messages \\
  NumMessagesSLTS & \hspace{0.2cm}- since last traige session \\ 
  NumMeetings & total num. of meetings \\
  NumMeetingsOrg & \hspace{0.2cm}- organized by the user \\
  TimeBusy & percent. of time user is busy \\
  TimeFree & percent. of time user is free \\
  TimeTentative & percent. of time user is tentative \\
  TimeOOO & percent. of time user is out of office \\
  \hline
  \rowcolor[gray]{.9}\multicolumn{2}{l}{Time Features} \\
  \hline
  HourOfDay & Hour of the day of first read \\
  DayOfWeek &  Day of the week of first read  \\
  DayOfMonth &  Day of the month of first read  \\
  Month  &  Month of first read \\
\hline
\end{tabular}
\end{small}
\label{tab:features}
\vspace{-17px}
\end{table}
\vspace{-0.9mm}
\subsection{Experimental setup and results}
We randomly split our dataset into a training set containing 80\% of the emails and a test set containing 20\% of the emails. Then, we perform 5-fold cross validation using the training set to select hyperparameters for the LightGBM classifier \cite{Ke_nips17} (an implementation of gradient boosted trees). During training, we weight positive examples 10 times more than negative examples because about 14\% of the examples are positive.

We compare our model against a simple baseline, which predicts as positive every message that has been marked as unread or flagged. Experiment 1 in Table \ref{predictiontable} shows that using the strongest deferral signals available to predict deferral results in surprisingly low precision and F1, suggesting that users often do not go back to messages they explicitly mark. We also see that our model is able to double precision and increase F1 by over 50\% compared to the baseline, showing that many additional signals are necessary to significantly improve prediction. For insight, we look at the features that have the highest Gini importance (Mean Decrease in Impurity )\cite{Ke_nips17} as output by the LightGBM classifier and find that the length of the body and the number of unhandled messages are the two most important features. This result is not surprising. The importance of the length of the email is bolstered by the conclusions from our qualitative study, which showed that users consider how long responding to an email will take when deciding whether to respond or not. And the importance of the number of unhandled messages is intuitive, since users may tend to defer more emails if they have more emails to attend to. 
\vspace{-0.5mm}
\begin{table}
\caption{Experiment 1: predicting whether a message marked as unread or flagged will contain a strong action in a later session %(Experiment 1), our model improves F1 by over 50\% compared to the baseline of predicting every email marked as unread or flagged as positive. 
Experiment 2: limiting the prediction to only those messages that contain strong actions %(Experiment 2), our model is only able to improve F1 by 12\%. 
Experiment 3: expanding the sample of messages from Experiment 2 to include messages not marked as unread and not flagged. Precision, and recall are denoted by P and R respectively.
%(Experiment 3), our model is able to improve F1 by almost 300\%. 
}
\begin{small}
\begin{tabular}{ l | l l l | l l l | l l l}
  
  \hline
  & \multicolumn{3}{c}{Experiment 1} & \multicolumn{3}{c}{Experiment 2} & \multicolumn{3}{c}{Experiment 3} \\
  \hline
   & P & R & F1 & P & R & F1 & P & R & F1\\
  \hline
 Our Model & 0.25 & 0.65 & 0.36 & 0.51 & 0.91 & 0.66 & 0.25 & 0.95 & 0.40\\
 Baseline  & 0.14 & 1.00 & 0.24 & 0.41 & 1.00 & 0.58 & 0.41 & 0.06 & 0.10\\ 
\hline
\end{tabular}
\end{small}
\label{predictiontable}
\vspace{-15px}
\end{table}

\begin{comment}
\hline\hline
  Experiment 2 & Precision & Recall & F1\\
\hline
  Our Model  	 & 0.513 & 0.908 & 0.656 \\
Baseline  		 & 0.410 & 1.0 & 0.582 \\
\hline\hline
  Experiment 3 & Precision & Recall & F1\\
\hline
  Our Model  	 & 0.253 & 0.947 & 0.400 \\  
  Baseline  		 & 0.410 & 0.059 &  0.103\\  
\end{comment}

\subsection{Predicting Deferral or Strong Action?}
\label{deferral}
We have shown that our model can reasonably make a compound two-part prediction: whether a user will return to an email \emph{and} perform a strong action on it. However, it is possible that our model is simply predicting whether a user will perform a strong action, since we excluded those emails with strong actions in the same session. We next investigate whether the act of deferral can be separated out from the act of replying or forwarding. 

We build a model to predict whether a user will return to an email after first read \emph{given} that they have performed a strong action on it at any time. In other words, we consider only those emails with strong actions, and predict those whose strong actions occur only in sessions later than the initial read session. By only considering emails with strong actions, we reduce the size of our dataset to 3339 messages. Needless to say that this prediction problem cannot be applied in practice, because we are trying to predict deferral after it has already happened. 
For this prediction problem, we exclude from the features the counts of strong actions, because these directly indicate the label. %We exclude the entire feature set because other actions in the session may heavily correlate with the strong actions. 
We set up our experiment exactly as before. Experiment 2 in Table \ref{predictiontable} shows that our model is only able to increase F1 over the baseline by about 12\%. We conclude that \emph{given} an email with a strong action, a mark as unread or flag signal is a strong signal that the email has been deferred, so the additional signals in our model are not that useful. However, note that our dataset only contains those messages that have been marked as unread or flagged (we made this restriction earlier for the practical prediction task). This restriction unfairly causes the baseline to have perfect recall, by definition.

Experiment 3 in Table \ref{predictiontable} shows that when we remove this restriction so that our dataset also includes messages not flagged and not marked as unread, our model is able to increase F1 over the baseline by almost 300\%, which suggests that our model is able predict whether a user will return to an email \emph{given} that the email contains a strong action. This result shows that our improvement in Experiment 1 is not just simple reply prediction.

\section{Discussion \& Design Implications}
\label{sec-discussions}
An overarching theme that emerged during our qualitative study was that while there has been a significant shift towards the centrality of email clients for task management, effective inbox management is still one of the main challenges facing enterprise workers who constantly deal with email overload. One main reason behind this observation is that the design of email clients has not kept up with the management scenarios it is expected to support. We identified three main areas to support users during their daily inbox management activities: (1) during triage (2) at the time of deferral and (3) while handling deferred emails.

When triaging, many of our participants try to use some of the available functionalities such as conditional formatting of the emails based on the sender, presence of explicit mention, etc. to facilitate the process of identifying emails that require attention. Automatic identification of emails that require attention, are urgent, or are likely to be deferred may have a great impact on managing email overload. 
%As well, while some of the participants had dedicated triage sessions to in the morning or at the end of the day to go through all of the recently delivered or pending items, they were only attending to the most important emails during the day.

Support for deferring emails was often requested by our participants. 
%While our prediction task was focused on emails that were read and then deferred, 
Our participants often commented on scenarios where emails can be automatically ``snoozed'' out of sight and resurfaced later. Many participants also described flags for responses and tasks, being able to specify the amount of time needed to handle emails and support for scheduling a time to revisit emails.

Finally, many of our participants expressed their fear of forgetting about their deferred emails and the need to use external tools due to the lack of client's support for effective task management. Reminders and notifications were envisioned as the most effective way to surface deferred emails. Integrating planners, task extractors, TODO list-generators, and automated support for suggesting the best times to handle deferred emails were also among the most popular functionalities for an intelligent email client. 
%[P4]: Another case is all the emails I receive when I have so many back to back meetings. Or when I have a big chunk of emails. So I have to go back to them when I'm done with the meetings. It would be nice if there is an email that is important or requires me to do something to pop up somehow so I notice it.
%So getting a notification as soon as the email is received won't be useful when I'm in meetings. Because I wouldn't be looking at it anyways. But having something like Snooze until meetings are finished would be very useful.
%So Outlook could tell me these are the major things you missed since the last time you checked your email.
%\aha{let's talk about email clients in general and not just outlook}
%\end{document}  % This is where a 'short' article might terminate

While prior work (e.g. \cite{Siu:2006}) had identified \textit{deferral} as a common email flow handling strategy, through in lab studies, to the best of our knowledge this work is the first attempt to formalize \textit{deferral} and how it can be inferred from logs of users actions. 
While we do replicate the previous design recommendations for enhancing the UI of email clients to leverage task views, scheduling tools, mechanisms for highlighting emails based on different metadata or means for annotating emails, our quantitative findings, based on the logs of thousands of users, highlighted different characteristics of messages (e.g. \#recipients, \#actionRequests or \#replyRequests) as well as deferral strategies (i.e. Flag or MarkAsUnread) that can be used to predict whether or not a read message is deferred. A direct implication of such prediction is designing email clients and intelligent assistants that help users with getting back to their pending messages. 

%\begin{acks}
%  The authors would like to thank ...
%\end{acks}

\section{Related Work}
\label{sec-relatedwork}

%Email is one of the most important means of online communication. As the volume of email grows, challenges related to email management and retrieval increase \cite{Castro}. 
Previous work has studied several aspects of how people interact with email and how to assist them with email management. 
Two lines of prior work are especially relevant, one on email management and organization and the other on large-scale log analysis of email interaction. We cover both of them below. 
\vspace{-0.3mm}
\paragraph*{Email organization and management}

One line of work on email management focused on understanding activities and workflows in email.
\citet{Venolia2001} identified five major activities surrounding how people use email. In particular, they highlighted two activities: keeping up with the flow of incoming messages, and triaging existing messages and discussed how mail clients could better support these activities.  \citet{Siu:2006} studied email use in the context of everyday work practices.
They examined how users interlace email with their day-to-day, ongoing work processes. They
demonstrated that subjects use email as a tool for managing moment-to-moment attention and task focus, and built on top of the work by ~\citet{Venolia2001} to propose a model of this workflow.

Much of the early research on email focused on how people organized and managed their email. \citet{Mackay:1988:MJC:62266.62293} described a set of interviews that focused on understanding the way professional workers use email. \citet{Whittaker}  proposed the concept of email overload to describe the usage of emails beyond communication needs, such as task management and personal archiving. They identified common strategies for handling email overload such as filing, searching, and cleaning. \citet{Fisher} found similar results in their study of mailboxes at a large tech company. \citet{Gwizdka} identified two additional email management practices, Cleaners and Keepers,
based on clustering responses to a questionnaire about email practices. \citet{Grevet} revisited these previous findings with a qualitative study of Gmail users and found that email overload was still prevalent in both work and personal settings.

\citet{Grbovic} showed that, with the increase of email messages over time, users do not use folders and argue that search is an increasingly important alternative to human-generated folders and tags. Several studies have focused on developing effective search systems for email~\cite{Dumais2003,Ramarao:2016}. \citet{Dumais2003} found that email was the most commonly retrieved source of personal information (e.g. files, web history, emails, etc.). 
%They also showed that people preferred to sort the results by date most often even when the default was best-match ranking.
Horvitz et al ~\cite{horvitz2005balancing} described experiments with bounded deferral, a method aimed at reducing the disruptiveness of incoming messages and alerts in return for bounded delays in receiving information. They showed that bounded deferral policies could help with balancing awareness about potentially urgent messages with the cost of interruption. 

Earlier work also focused on studying how people triage their email. \citet{Neustaedter:2005} performed a set of interviews and surveys to understand how people handle email during triage and what email do people decide to handle first. \citet{pierce2010triage} also studied triage but they focused on mobile use. They showed that triage is a more dominant activity on mobile and that users often triage on mobile and defer their action until they reach a desktop. Perhaps the most relevant to our work in this line is the work of \citet{Wainer:2011}. They studied how top-level cues, including message importance, organization utility, subject line specificity, curiosity, workload and personal utility, influence attention to email. 
%That is, why people attend to some emails and not others based on inbox-level cues about message content. 
%They found several factors including message importance, organization utility, subject line specificity, curiosity, workload and personal utility. The study was based on think-aloud sessions with five participants where participants accessed their work email account and were asked to verbalize their thoughts while selecting emails they would like to read immediately. 

Our work is similar to this line of research in that we also study aspects of email management and organization. We examine in greater detail one aspect of email management focused on deferred action on emails and provide in-depth qualitative and quantitative studies to characterize and support users with deferral. 
\paragraph*{Log-based analysis of email interaction}
Large-scale log analysis has been extensively used in the literature to study different aspects of email interactions.  \citet{Kalman} conducted a study of email management strategies on thousands of users over a period of 8 months using a popular email web client add-in. They showed that people use a wide variety of strategies to manage their emails, many more than had been identified in earlier studies.
\vspace{-0.3mm}
Other work focused on using large-scale log analysis to study re-finding in email. \citet{Elsweiler} studied several email interactions such as sorting, changing views,  searching, selecting messages and opening folders.  Their work revealed strong relationships between
difficulty in re-finding emails and the time lapsed since a message was read, remembering when the sought-after email was sent and remembering other recipients of the email.
%, the experience of the user and the user's filing strategy. 
\citet{Whittaker2} carried out a large-scale study of users using a web-based email client. They investigated different re-finding strategies. They found out that some users create and use complicated folder structures to use them for email retrieval with no improvement in retrieval success. 
%However, they noted that this did not result in improving retrieval success compared to users who rely on search.  
\citet{Dumais2003} showed that re-finding previously seen information is a frequent activity that goes beyond email. However, they showed that email is by far the most common type of information that people re-find in a desktop search application.
%, and that more than half of the items re-found using search are more than a month old.

Log-based studies have also been used to study email search. \citet{Ai} examined the actions that people perform on emails after searches and compared re-finding in email search with web search. \citet{Narang} also examined the activities performed on messages following searches, and how this related to the characteristics of people's mailboxes and email organization strategies. 
%They found that people with larger mailboxes search more, and people who organize less tend to search more.

More recently, \citet{Castro} studied what actions the users might conduct on received messages by analyzing the actions of a large number of users in Yahoo!~Mail. They found out that the most frequent actions are typically read, reply, delete and a sub-type of delete, delete-without-read. Yang et al. \cite{Yang:2017} studied email reply behavior in enterprise settings. They characterized the influence of various factors such as email content and metadata, historical interaction features and temporal features on email reply behavior. We also develop models
to predict whether a recipient will reply to an email and how long it will take to do so

\citet{Alrashed:2018:LEM:3176349.3176398} studied the lifetime of email messages with a focus on revisiting patterns. They showed that some emails are never revisited, while others are revisited for multiple times. They also showed that some users revisit messages for a variety of reasons including taking an action or for finding location information.

This line of work is related to our study since we also use large-scale log analysis to study email interactions. We build on top of previous work on understanding email triage and re-finding but focus on the act of email deferral. We also augment the log-based analysis with a qualitative study and use the findings of both studies to define a prediction task to help users get back to deferred emails more efficiently.  

\vspace{-0.3mm}
\section{Conclusions}
\label{sec-conclusions}
We sought to understand one important aspect of email triage where people defer taking action on emails after seeing it for the first time. We employed a mixed-methods approach where we combined qualitative analysis, via interviews, and quantitative analysis, via a large-scale log study, to develop a better understanding of why people defer emails, what strategies they employ to do so and how we can provide better support for deferring and getting back to deferred emails.

We found that users are more likely to defer emails that require responding or processing information in an attachment. We also observed that deferred message share common characteristics related to the importance of the sender, the number of recipients and the content of the message. 
% However some of these characteristics were also observed for messages that received an immediate strong action and where not deferred. 
Further, the decision to defer a message depends not only on the message itself but also on other contextual information such as the current workload of the user. Our log analysis revealed several insights about deferral strategies. For example, we showed that Flag and MarkAsUnread are more likely to be observed for deferred than non-deferred  messages but are only observed for a limited portion of deferred messages. We used these insights to define a prediction task to assess the feasibility of using machine learning to assist users with their deferral workflows.

Understanding email deferral could have implications on understanding how people interact with their email and designing email clients and intelligent agents to help people with managing and organizing their messages. Our future work will aim to explore applications and user experiences that would better support email deferral.

\bibliographystyle{ACM-Reference-Format} 
%\bibliography{strings-short,sample-bibliography} 
\bibliography{shortref}

\end{document}